\documentclass[a4paper,fleqn]{cas-dc}

\usepackage{multirow}

\def\tsc#1{\csdef{#1}{\textsc{\lowercase{#1}}\xspace}}
\tsc{WGM}
\tsc{QE}
\tsc{EP}
\tsc{PMS}
\tsc{BEC}
\tsc{DE}

\usepackage[dvipsnames]{xcolor}
\usepackage{algorithm}
\usepackage{algorithmic}
\usepackage{caption}
\usepackage{changepage}
\usepackage{tabularx}
\usepackage{array}
\usepackage[square,sort,comma,numbers]{natbib}
\setcitestyle{square}
\begin{document}
\let\WriteBookmarks\relax
\def\floatpagepagefraction{1}
\def\textpagefraction{.001}

\shorttitle{EdgeAISim: A Toolkit for Simulation and Modelling of AI Models in Edge Computing Environments}

\shortauthors{Aadharsh Roshan Nandhakumar et al.}

\title [mode = title]{EdgeAISim: A Toolkit for Simulation and Modelling of AI Models in Edge Computing Environments}                      

\author[1,4]{Aadharsh Roshan Nandhakumar}[orcid=0000-0003-0086-8478]
\affiliation[1]{organization={Indian Institute of Information Technology Allahabad India.}}
\ead{aadh2001@gmail.com}

\author[1,4]{Ayush Baranwal}
\ead{baranwalayush25@gmail.com}

\author[2,4]{Priyanshukumar Choudhary}
\affiliation[2]{organization={National Institute of Technology, Rourkela, India.}}
\ead{priyanschoudhary100@gmail.com}

\author[3,4]{Muhammed Golec} 
\affiliation[3]{organization={Abdullah Gul University, Kayseri, Turkey.}}
\affiliation[4]{organization={School of Electronic Engineering and Computer Science, Queen Mary University of London, London, UK}}
\ead{m.golec@qmul.ac.uk}
\cormark[1]

\author[4]{Sukhpal Singh Gill}[orcid=0000-0002-3913-0369] 
\ead{s.s.gill@qmul.ac.uk}

\cortext[cor1]{Correspondence to: School of Electronic Engineering and Computer Science, Queen Mary University of London, London, E1 4NS, UK.}

\begin{abstract} 
To \textcolor{black}{meet next-generation Internet of Things (IoT) application demands, edge computing moves processing power and storage closer to the network edge to minimise latency and bandwidth utilisation. Edge computing is becoming increasingly popular as a result of these benefits, but it comes with challenges such as managing resources efficiently. Researchers are utilising Artificial Intelligence (AI) models to solve the challenge of resource management in edge computing systems. However, existing simulation tools are only concerned with typical resource management policies, not the adoption and implementation of AI models for resource management, especially. Consequently, researchers continue to face significant challenges, making it hard and time-consuming to use AI models when designing novel resource management policies for edge computing with existing simulation tools. To overcome these issues, we propose a lightweight Python-based toolkit called EdgeAISim for the simulation and modelling of AI models for designing resource management policies in edge computing environments. In EdgeAISim, we extended the basic components of the EdgeSimPy framework and developed new AI-based simulation models for task scheduling, energy management, service migration, network flow scheduling, and mobility support for edge computing environments. In EdgeAISim, we have utilised advanced AI models such as Multi-Armed Bandit with Upper Confidence Bound, Deep Q-Networks, Deep Q-Networks with Graphical Neural Network, and Actor-Critic Network to optimize power usage while efficiently managing task migration within the edge computing environment. The performance of these proposed models of EdgeAISim is compared with the baseline, which uses a worst-fit algorithm-based resource management policy in different settings. Experimental results indicate that EdgeAISim exhibits a substantial reduction in power consumption, highlighting the compelling success of power optimization strategies in EdgeAISim. The development of EdgeAISim represents a promising step towards sustainable edge computing, providing eco-friendly and energy-efficient solutions that facilitate efficient task management in edge environments for different large-scale scenarios.}
\end{abstract}

% % Use if graphical abstract is present
% \centerline{\includegraphics[width=0.9\linewidth]{Figure2.pdf}} 

% % Research highlights
% \begin{highlights}
% \item EdgeAISim offers simulations of AI models for resource management in edge computing.
% \item EdgeAISim offers task scheduling, service migration, and network flow scheduling.
% \item EdgeAISim provides energy management and mobility support for edge computing.
% \item EdgeAISim outperforms the worst-fit algorithm-based resource management baseline.
% \item EdgeAISim represents a promising step towards sustainable edge computing.
% \end{highlights}

% Keywords

\maketitle

\section{Introduction}

\textcolor{black}{
In the contemporary digital landscape, edge computing has emerged as a seminal approach that significantly reshapes data handling, processing, and analysis \cite{cruz2022edge}. It represents a pivotal departure from the conventional centralized data processing infrastructure of distant cloud data centers  \cite{nabavi2022tractor}. Edge computing involves a decentralized approach to computing, focusing on processing and analyzing data closer to where it originates, near the network's edge, rather than relying on central cloud data centers \cite{murshed2021machine}. The decentralized approach of edge computing plays a pivotal role in managing the substantial data volumes generated by the Internet of Things (IoT) \cite{hua2023edge}. In IoT, edge computing efficiently handles data from myriad devices. Its significance extends to diverse applications such as augmented reality/virtual reality (AR/VR), autonomous vehicles, smart cities, telecommunications, healthcare, and video surveillance, all of which necessitate rapid data processing \cite{edgeai}.}

\subsection{Challenges}
\par
\textcolor{black}{The rising demand for edge computing is propelled by the ever-increasing volume of data in the digital age, primarily attributed to the IoT \cite{iftikhar2022ai}. The rapid proliferation of edge computing has ushered in a new era of decentralized data processing, promising reduced latency, enhanced privacy, and improved efficiency \cite{du2023computation}. This transition, however, presents significant challenges, with power consumption taking the forefront \cite{jiang2022joint}. It becomes imperative to optimize energy usage for sustainability, cost reduction, and resource allocation efficiency for modern edge computing systems  \cite{nabavi2023seagull}. In summary, the imperatives of sustainability, cost-effectiveness, and efficient resource allocation become increasingly pronounced as the popularity of edge computing continues to soar \cite{gill2022ai}. Effective resource management is vital for maintaining consistent performance and extending the lifespan of edge servers while mitigating thermal constraints \cite{aslanpour2021serverless}. Researchers are utilising Artificial Intelligence (AI) models to solve the above-mentioned challenges of resource management in edge computing systems  \cite{ghafouri2022mobile}. However, existing simulation tools are only concerned with typical resource management policies, not the adoption and implementation of AI models for resource management, especially  \cite{aslanpour2020performance}. Consequently, researchers continue to face significant challenges, making it hard and time-consuming to use AI models when designing novel resource management policies for edge computing with existing simulation tools \cite{MAHMUD2022111351} \cite{EdgeCloudSim} \cite{gupta2017ifogsim} \cite{souza2023edgesimpy}.}

\subsection{Existing Solutions}
\textcolor{black}{EdgeSimPy \cite{souza2023edgesimpy} is a Python-based framework tailored for modeling and simulating resource management policies within edge computing ecosystems and it solves the issues of existing simulators \cite{MAHMUD2022111351} \cite{EdgeCloudSim} \cite{gupta2017ifogsim}. Distinguished by its modular architecture, EdgeSimPy encompasses functional abstractions for various components, including edge servers, network devices, and applications. This versatile tool empowers users to explore and optimize resource allocation strategies, enhancing the efficiency and performance of edge computing environments. In this section, we highlights the limitations of existing simulators, particularly EdgeSimPy and elucidates how our proposed toolkit, EdgeAISim overcomes these deficiencies through the integration of AI techniques tailored for edge computing environments. }

 \subsubsection{Limited Realism and Adaptability}

\textcolor{black}{
Traditional simulators face difficulties representing the dynamic, heterogeneous edge environment realistically and adapting to diverse scenarios.}
\par \textcolor{black}{EdgeAISim, powered by AI, self-optimizes for enhanced realism. Reinforcement learning equips it to dynamically adapt resource management, ensuring a more realistic and adaptable depiction of edge computing dynamics in evolving conditions.}

\subsubsection{Inefficient Resource Management}

\textcolor{black}{
Conventional simulators often use basic resource allocation methods, neglecting power optimization and task migration, resulting in suboptimal resource use.}

\par \textcolor{black}{ EdgeAISim, driven by AI, dynamically manages resources, reducing power consumption. It employs techniques like Multi-Armed Bandit and Actor-Critic Reinforcement Learning for intelligent resource allocation, ensuring efficiency.}

\subsubsection{Lack of Adaptation to Emerging AI Trends}

\textcolor{black}{
Existing simulators often lack the capacity to assess emerging AI technologies and advanced reinforcement learning in enhancing edge computing efficiency.}

\par \textcolor{black}{ Whereas, EdgeAISim positions itself at the forefront of research and development in edge computing. It can explore the synergy between AI and edge computing, enabling researchers to assess the potential benefits of cutting-edge AI techniques in real-world edge scenarios.}

\subsection{Our Contributions}
\textcolor{black}{This paper meticulously explores the vast spectrum of applications of edge computing, accentuating its pivotal role in addressing the torrential influx of data generated during the IoT era. Furthermore, we introduce EdgeAISim, an advanced framework that offers cost-effective and sustainable resource management for edge computing systems. By focusing on sustainability, cost-effectiveness, and resource optimization, EdgeAISim strives to address a critical challenge in the evolving realm of edge computing.} The \textbf{\textit{main contributions}} of this work are:
\begin{itemize}
\item We proposed EdgeAISim, a Python-based framework that addresses the challenges of task migration and resource management in dynamic edge computing environments. Leveraging advanced algorithms and simulations, EdgeAISim enables seamless task migration and workload balancing across edge servers.
\item We integrated advanced AI models such as Multi-Armed Bandit with Upper Confidence Bound, Deep Q-Networks, Deep Q-Networks with Graphical Neural Network, and Actor-Critic Network to optimize power usage while efficiently managing task migration within the edge computing environments.
\item We compared the performance of these proposed models of EdgeAISim with the baseline, which uses a worst-fit algorithm-based resource management policy in different settings. Experimental results indicate that EdgeAISim exhibits a substantial reduction in power consumption, highlighting the compelling success of power optimization strategies in EdgeAISim. EdgeAISim minimizes power consumption, promoting sustainability, and reducing the environmental impact of energy-intensive infrastructures.
\end{itemize}

\subsection{Article Organization}
The rest of the paper is organised as follows: Section 2 presents related work. Section 3 discusses background, including definitions and concepts. Section 4 presents the architecture of EdgeAISim. Section 5 discusses the design and implementation of
EdgeAISim. Section 6 presents the experimental setup and results. Finally, Section 7 concludes the paper and highlights future directions.

\section{Related Work}

The advent of edge computing has sparked the creation of numerous edge simulators. In this section, we provide an introduction to simulation tools dedicated to edge computing and subsequently conduct a comparative analysis, focusing on the distinct features and contributions of EdgeAISim as compared to these existing solutions. By evaluating capabilities of the EdgeAISim, we aim to highlight its potential advantages and novel contributions to the field of edge simulation. Calheiros et al. \cite{goyal2012cloudsim} developed CloudSim, a versatile simulator for cloud computing infrastructure. CloudSim aids researchers in conducting comprehensive studies on resource allocation, task scheduling, and energy consumption. It enables in-depth simulations, enhancing the development of cloud-based applications and the evaluation of cloud management strategies. Further, Sonmez et al. \cite{EdgeCloudSim} extends the CloudSim \cite{goyal2012cloudsim} and address the limitations of cloud and network simulators in modeling edge environments. While cloud simulators like CloudSim lack user mobility and wireless support, network simulators may not cover edge servers and users. They introduce EdgeCloudSim \cite{EdgeCloudSim}, a simulator enabling user mobility, edge device power modeling, and network management for edge computing scenario prototyping. Moreover, Mahmud et al. \cite{MAHMUD2022111351} introduced iFogSim2, an enhanced version of iFogSim \cite{gupta2017ifogsim}. This toolkit emphasizes mobility, clustering, and microservice management in edge and fog computing environments. Researchers can explore the impact of mobility, clustering strategies, and microservice management in these dynamic scenarios. iFogSim2 empowers researchers to optimize edge and fog computing applications for improved mobility, clustering, and microservice utilization.

Zeng et al. \cite{ZENG201793} introduced IOTSim, a dedicated simulator for analyzing IoT applications. This tool enables researchers to study the behavior and performance of IoT applications comprehensively. IOTSim facilitates resource management, communication protocols, and data processing evaluations in IoT scenarios, empowering researchers to optimize IoT solutions and enhance IoT applications' efficiency. \textcolor{black}{ Further, Jha et al. \cite{IOTSim-Edge} tackle the complex Cloud-Edge-IoT ecosystem, involving resource allocation across diverse devices with various network and messaging protocols. To address the limitations of edge simulators for these protocols, they present IoTSim-Edge. This simulator enables experimentation with edge resource management, considering factors like energy use, application composition, user mobility, and communication protocols. } \textcolor{black}{Alwasel et al. \cite{IOTSim-Osmosis} propose IoTSim-Osmosis, a novel simulator designed for Osmotic Computing scenarios to focus on workload migration between cloud data centers and edge devices based on performance and security events. This simulator includes various models for data transmission, energy consumption, and application performance. The authors also demonstrate IoTSim-Osmosis in a case study, showcasing its ability to model policies optimizing performance, energy usage, and cost in Cloud-Edge scenarios.}

\textcolor{black}{Wang et al. \cite{SimEdgeIntel} introduces a versatile edge caching simulator offering key contributions. It facilitates quick mobile network setup, heterogeneous device simulation, and various scenarios. Its unique algorithm access framework simplifies integration. The simulator is highly flexible, supports custom models, and includes a learning-based caching algorithm for cloud collaborative intelligence, as demonstrated through performance evaluations. } \textcolor{black}{ Qayyum et al. \cite{Fognetsim++} critique existing edge simulators for their limitations in capturing network infrastructure nuances. They present FogNetSim++, an innovative simulator that addresses these issues by modeling power consumption, supporting various communication protocols, and simulating mobile user handovers. They demonstrate its effectiveness in a practical use case for designing edge-specific placement and scheduling policies.}
\par
\textcolor{black}{Lera et al. \cite{YAFS} present YAFS, an edge simulator focused on evaluating allocation decisions in composite applications within edge infrastructures. YAFS employs routing policies to oversee communication among application modules, enabling the allocation of application components to diverse edge devices. The simulator showcases applications, including dynamic scheduling, infrastructure resilience, and user mobility support. Recently, Souza et al. \cite{Edgesimpy} presents a Python-based framework called EdgeSimPy for modeling and simulating resource management policies within edge computing ecosystems. EdgeSimPy encompasses functional abstractions for various components, including edge servers, network devices, and applications.}

\subsection{Critical analysis}
Table \ref{table:1} shows the comparison of EdgeAISim with existing frameworks and simulators. \textcolor{black}{Our research paper stands out in the field due to its innovative approach, which specifically targets the optimization of power consumption in edge computing systems using AI models, especially reinforcement learning techniques. This breadth of capabilities of EdgeAISim makes it a versatile tool for modeling and evaluating various edge computing scenarios. In contrast, while other simulators like CloudSim \cite{goyal2012cloudsim} and EdgeCloudSim \cite{EdgeCloudSim} excel in specific areas such as task scheduling, they often lack the holistic approach that EdgeAISim provides. iFogSim2 \cite{MAHMUD2022111351}, while encompassing energy management and service migration, lacks AI and network flow scheduling. To harness the full potential of edge computing, it necessitates advanced simulation tools capable of addressing its unique challenges to simulate AI based energy-efficient resource management polices for edge computing systems, which is also missing in EdgeSimPy \cite{Edgesimpy}.  Furthermore, our proposed simulator, EdgeAISim has ability to simulate the full spectrum of edge computing functionalities positions it as a promising choice for researchers seeking to conduct in-depth assessments of edge computing systems and applications, making it stand out in the landscape of edge computing simulation tools.}

\begin{table*}
\centering
\caption{Summary of built-in features supported by existing simulators and EdgeAISim}

\begin{tabular}{|p{2.6 cm}|p{0.5cm}p{1.5 cm}p{1.8cm}p{1.4 cm}p{1.6cm}p{1.5 cm}p{1.2 cm}p{1.9 cm}|}

\hline
Simulator & AI & Task Scheduling & Energy Management & Service Migration & Maintenance Operation & RL Algorithms &   Mobility Support & Network flow Scheduling\\
\hline

CloudSim \cite{goyal2012cloudsim} &$\times$  & $\checkmark$ & $\checkmark$ & $\times$ & $\times$ &$\times$ & $\checkmark$ & $\times$  \\
EdgeCloudSim \cite{EdgeCloudSim} & $\times$ & $\checkmark$ & $\checkmark$& $\times$& $\times$ & $\times$ &$\checkmark$ & $\checkmark$   \\
IOTSim \cite{ZENG201793} & $\times$& $\checkmark$ & $\times$ & $\times$ & $\times$ & $\times$ &$\checkmark$ & $\checkmark$  \\
IFogSim2 \cite{MAHMUD2022111351}& $\times$& $\checkmark$ & $\checkmark$ & $\checkmark$& $\checkmark$ & $\times$ &$\checkmark$ & $\checkmark$  \\
SimEdgeIntel \cite{SimEdgeIntel} &$\times$  & $\times$ & $\times$ & $\checkmark$  & $\times$ &$\checkmark$ & $\checkmark$ & $\checkmark$  \\

IoTsim-Edge \cite{IOTSim-Edge} &$\times$  & $\checkmark$ & $\checkmark$ & $\times$  & $\times$ &$\times$ & $\checkmark$ & $\checkmark$  \\
IoTsim-Osmosis \cite{IOTSim-Osmosis} &$\times$  & $\checkmark$ & $\checkmark$ & $\times$  & $\times$ &$\times$ & $\checkmark$ & $\checkmark$  \\
FogNetSim++ \cite{Fognetsim++} &$\times$  & $\checkmark$ & $\checkmark$ & $\times$  & $\times$ &$\times$ & $\checkmark$ & $\times$  \\
YAFS \cite{YAFS} &$\times$  & $\checkmark$ & $\checkmark$ & $\times$ & $\times$ &$\times$ & $\checkmark$ & $\times$  \\
EdgeSimPy \cite{Edgesimpy} & $\times$ & $\checkmark$ & $\times$ & $\checkmark$ & $\checkmark$ &$\times$ & $\checkmark$ & $\checkmark$  \\
EdgeAISim (this paper) & $\checkmark$ & $\checkmark$ & $\checkmark$ & $\checkmark$ & $\checkmark$ &$\checkmark$ & $\checkmark$ & $\checkmark$  \\

\hline
\end{tabular}

\label{table:1}
\end{table*}

\section{\textcolor{black}{Background: Definitions and Concepts}}
This section discusses important definitions and concepts to understand this work.
\subsection{Edge Computing}

\begin{figure}[t]
    \centering
    \includegraphics[width=1\linewidth]{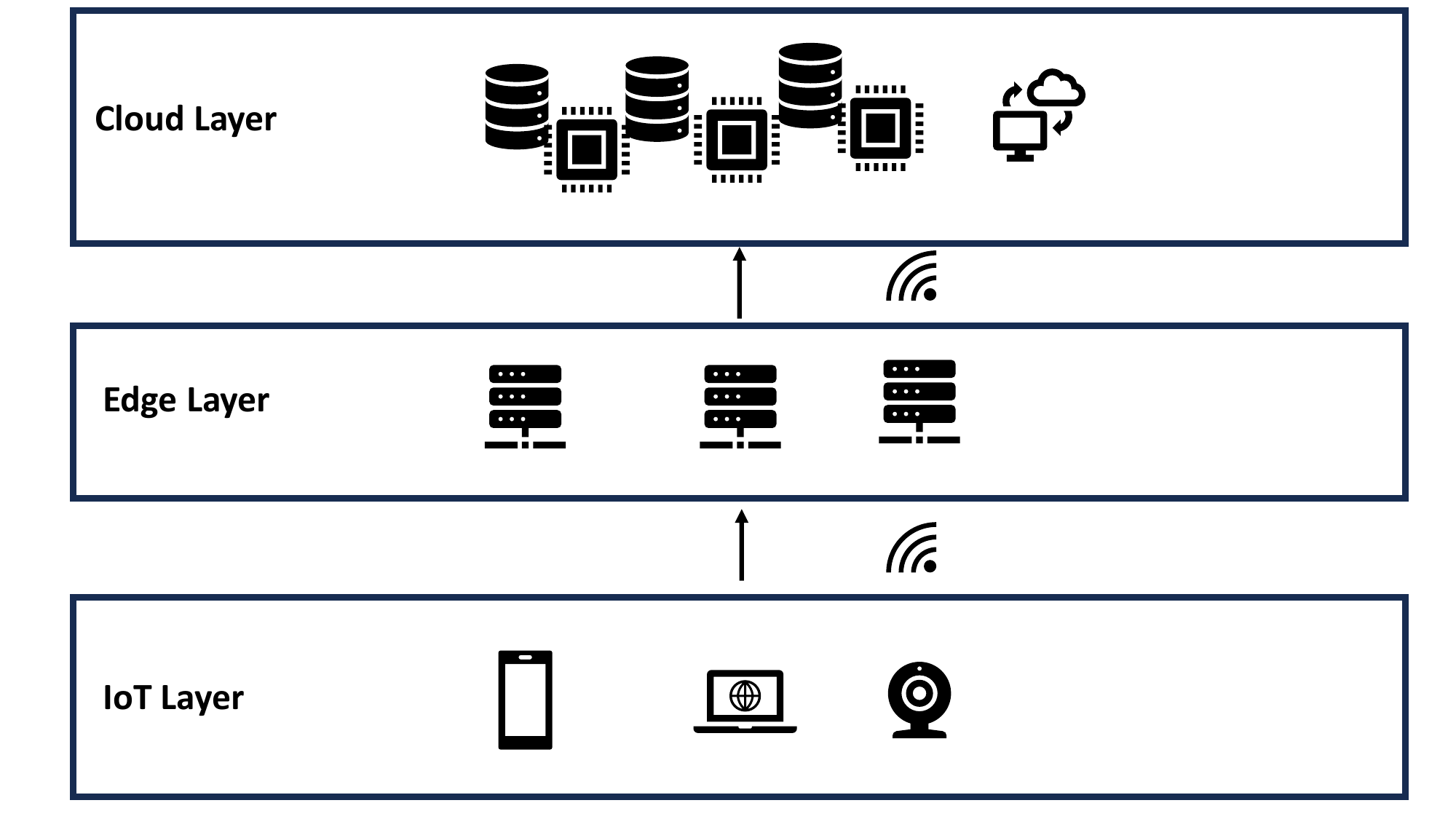}
    \caption{\textcolor{black}{Edge Computing Environments}}
    \label{fig:background}
\end{figure}

\textcolor{black}{
Figure \ref{fig:background} shows the basic architecture of edge computing, where it interacts with the IoT layer and the cloud computing layer. Edge computing is a computing model that places resources closer to the network edge, enabling faster data processing and real-time capabilities \cite{saleh2022trust}. It brings computing power and services nearer to data sources and end-users, reducing latency and improving application performance.}
\subsection{Problem Formulation: Task Migration in Edge Computing}
\textcolor{black}{
The task migration problem refers to the challenge of efficiently relocating and resuming the execution of computing tasks or processes from one computing resource, such as a server or a Virtual Machine (VM), to another while minimizing disruption and optimizing resource utilization \cite{zhang2019task}. In the realm of edge computing, the task migration challenge takes on heightened significance. Here, tasks frequently require dynamic relocation among edge devices for a range of purposes, including load distribution, fault resilience, energy conservation, and adapting to evolving resource requirements \cite{liang2021multi}. Proficient task migration strategies hold paramount importance within edge computing, as they serve to curtail service interruptions, mitigate data transfer overhead across potentially constrained network links, and enhance the efficient use of resources \cite{tang2021task}. These strategies are indispensable for sustaining continuous and responsive edge services, particularly for applications demanding real-time performance and low-latency responsiveness.}

 \begin{figure*}[htb]
\centerline{\includegraphics[width=0.9\linewidth]{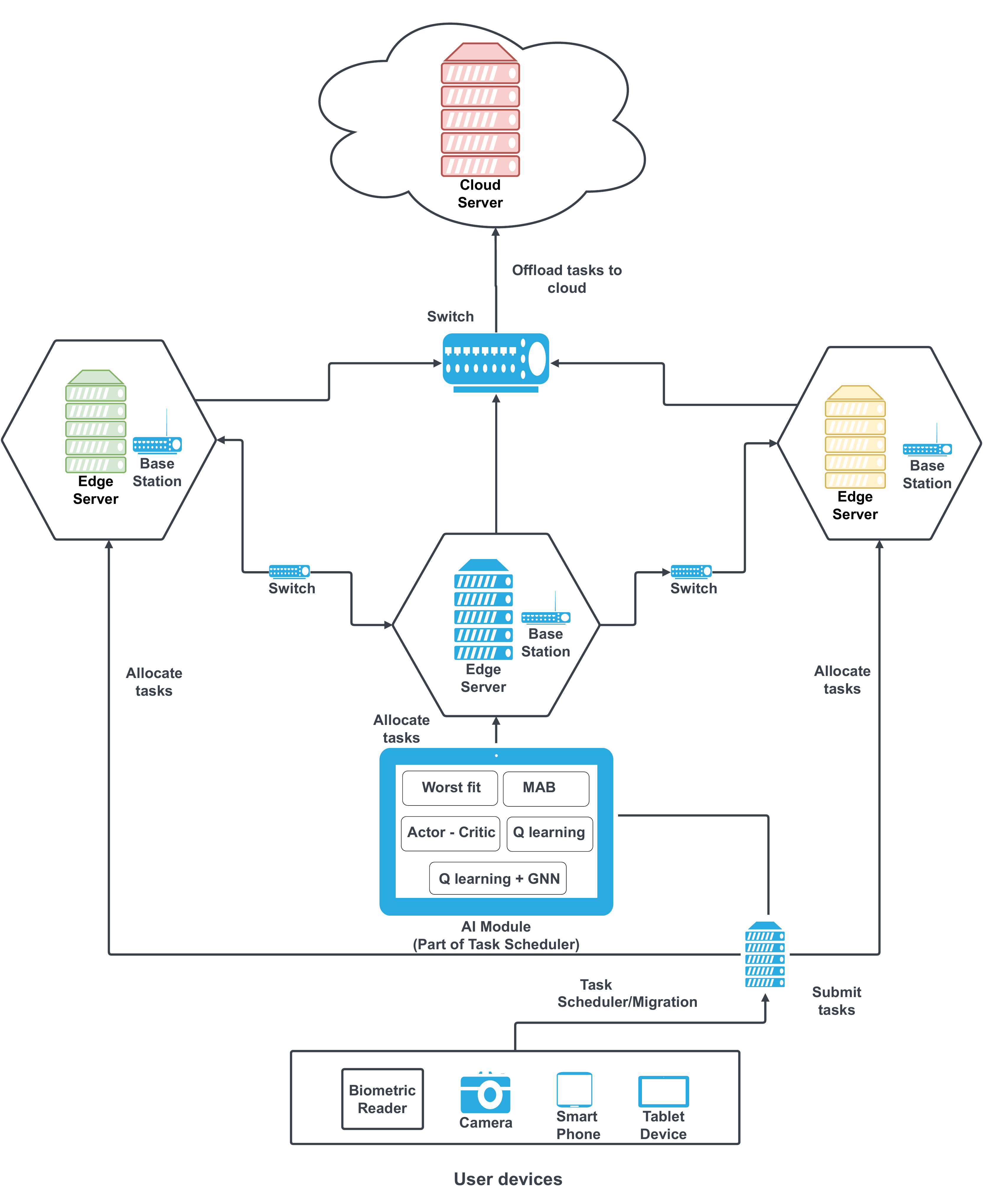}} 
\caption{\textcolor{black}{EdgeAISim Architecture }}
\label{main}
\end{figure*}

\section{\textcolor{black}{EdgeAISim Architecture}}
\textcolor{black}{Figure \ref{main} shows the system architecture of EdgeAISim. In EdgeAISim, we extended the basic components of the EdgeSimPy framework \cite{Edgesimpy} and developed new AI-based simulation models for task scheduling, energy management, service migration, network flow scheduling, and mobility support for edge computing environments. A basic edge cloud network consists of the following components:}

\subsection{Base stations}

 Base stations provide network connectivity to mobile computing devices within their coverage area. The entire map is divided into cells, where each base station assumes coverage of each cell. A mobile computing device anywhere in the cell is assumed to have equal connectivity to the cell's base station.

 \subsection{Network Switches}

 Network switches are used to provide connections, typically wired, between base stations and edge servers. Task migrations are typically modeled as network flows, whose duration is determined by bandwidth scheduling. The Max-Min fairness algorithm \cite{hosaagrahara2008max} is used for bandwidth scheduling in network switches.

 \subsection{Modelling of Resources}

 Edge servers are used to host services. Power consumption is modeled using three built-in power consumption models: LinearPowerModel, QuadraticPowerModel, and CubicPowerModel, which are dependent on 3 parameters: CPU, RAM, and hard disk utilization. \textcolor{black}{ In a linear power model, the parameters are dependent upon each other in a degree 1 polynomial equation. In a quadratic power model; the parameters are dependent upon each other in a degree 2 polynomial equation, whereas in a cubic power model, the parameters are dependent upon each other in a degree 3 polynomial equation.} Edge servers run virtualized containers, enabling them to run multiple services at the same time. 

 \subsection{Users}

 Users are the consumers of the services hosted on the edge Servers. Users move according to a defined mobility model, changing their access base station for the consumption of services present on the edge server.

 \subsection{\textcolor{black}{Modelling of Tasks}}

 \textcolor{black}{Tasks are modelled as applications or services. These applications or services have definite resource requirements, which are CPU demand and memory demand. When these are allocated to a specific edge server, the edge server will begin consuming the resources and, consequently, the change is reflected in its power consumption.}

 \subsection{\textcolor{black}{Modelling of AI Models}}

 \textcolor{black}{The AI module consists of Reinforcement Learning (RL)-based task migration algorithms for the effective allocation of tasks, which is explained in the Section \ref{sec:Design}.}

\begin{figure*}[htb]
\centerline{\includegraphics[width=1\linewidth]{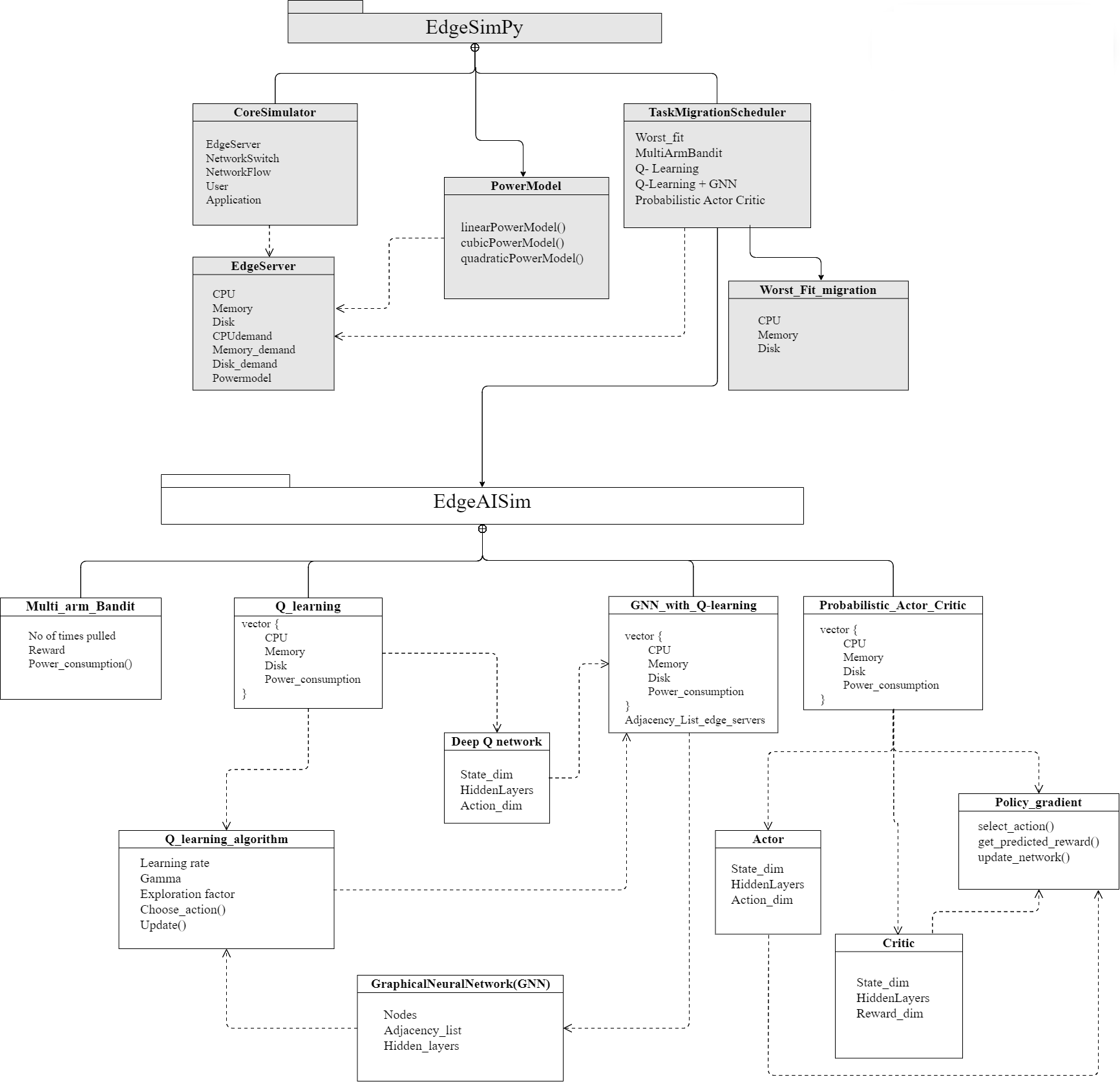}} 
\caption{\textcolor{black}{Fundamental Classes of EdgeAISim}}
\label{fig:fig3}
\end{figure*}

 \section{\textcolor{black}{Design and Implementation of EdgeAISim}\label{sec:Design}}
This section discusses the design and implementation of EdgeAISim. 
\subsection{\textcolor{black}{Design Description}}
The fundamental classes of EdgeAISim are shown in Figure \ref{fig:fig3}. The following are the main classes of EdgeAISim:

 \subsubsection{\textcolor{black}{Core Simulator}}
 
 \textcolor{black}{This class serves as the central orchestrator within the EdgeAISim system. Its primary role is to facilitate the seamless coordination and operation of various integral components, including network switches, edge servers, network flows, users, and more. One of its crucial functions is to instantiate instances of these component classes, creating individual entities representing network switches, edge servers, network flows, and users. Once these entities are created, this coordinating class assigns each of them to their designated positions and roles within the system. }

 \subsubsection{\textcolor{black}{PowerModel: Modelling of Energy Consumption}}

 \textcolor{black}{This class is responsible for the practical implementation of several power models within the context of edge server and potentially other system components as well. These power models are essential tools for simulating and estimating power consumption. In the context of edge servers, power consumption is a crucial factor, as they often operate in resource-constrained and energy-sensitive environments. The class takes on the task of translating theoretical power models into functional code, allowing the simulation to accurately predict how much power an Edge server (or other components) will consume under different conditions.}

\subsubsection{\textcolor{black}{Task Migration Scheduler}}

\textcolor{black}{This class serves as a pivotal component within the edge computing system, responsible for the allocation of tasks to different edge servers. Its primary role is to make intelligent decisions about how to distribute tasks efficiently among available servers. The AI module, consisting of varying RL algorithms, is implemented by extending this class. Section \ref{sec:AIModels} discusses the modelling and implementation of AI models for task migration. }

\subsection{Modelling and Implementation of AI Models for Task Migration}\label{sec:AIModels}

A task may be added to a migration queue due to Quality of Service (QoS) requirements, and the migration function called, which defines the migration algorithm. In this work, we have implemented a baseline resource management policy using a worst-fit algorithm \cite{lai2020cost, xu2022dynamic}. Further, four advanced AI models (RL algorithms) such as Multi-Armed Bandit with Upper Confidence Bound, Deep Q-Networks, Deep Q-Networks with Graphical Neural Network, and Actor-Critic Network are utilized to optimize power usage while efficiently managing task migration within the edge computing environment. Each time the migration algorithm is called is modeled as one timestep. At each timestep, the inverse of the power consumption of each server is added and the total is used as our reward. The RL algorithm attempts to maximize our reward, and hence minimize our power consumption.

 \subsubsection{Baseline Model: Worst-Fit Migration Algorithm}
 We have considered a worst-fit method \cite{lai2020cost, xu2022dynamic} to develop a baseline policy for resource management. An algorithmic representation of Worst-Fit Algorithm is shown in Algorithm~\ref{alg:WorstFit}.  This is a simple migration algorithm, where we sort the edge server by the difference between the available CPU and CPU demand in a decreasing order. Each service is migrated into the first server which has the available system requirements.

 \begin{algorithm}
\caption{Worst-Fit Algorithm for Task Migrations in Edge-Cloud Computing}
\begin{algorithmic}
\STATE \textbf{Input:} Set of tasks $\mathcal{T}$, Set of edge servers $\mathcal{E}$, Migration threshold $M_{\text{threshold}}$

\STATE \textbf{Initialize:} Mapping of tasks to edge servers, $Mapping \leftarrow \{\}$

\FOR{each task $t_i \in \mathcal{T}$}
    \STATE \textbf{Find Candidate Edge Servers:}
    \STATE $CandidateServers \leftarrow$ Find all edge servers $e_j \in \mathcal{E}$ with sufficient resources to accommodate $t_i$
        \STATE \textbf{Select Worst-Fit Edge Server:}
        \STATE Sort $CandidateServers$ in decreasing order of available resources
        \STATE $e_{\text{worst}} \leftarrow$ First server with the available resources among $CandidateServers$
        
            \STATE Offload task $t_i$ to edge server $e_{\text{worst}}$
            \STATE $Mapping[t_i] \leftarrow e_{\text{worst}}$

\ENDFOR

\STATE \textbf{Output:} Final mapping of tasks to edge servers $Mapping$
\end{algorithmic}
\label{alg:WorstFit}
\end{algorithm}

 \subsubsection{\textcolor{black}{Proposed Model: Multi-Armed Bandit with Upper Confidence Bound}}

 \textit{Multi-Arm Bandit} refers to a decision-making problem where an agent must choose between multiple options (arms) with unknown rewards, aiming to maximize cumulative rewards by balancing exploration and exploitation \cite{MultiArmBandit}. \textit{Upper Confidence Bound (UCB)} is a popular algorithm used in the Multi-Armed Bandit problem to balance exploration and exploitation \cite{UpperConfidenceBound}. An algorithmic representation of Upper Confidence Bound (UCB) Algorithm for Multi-Arm Bandit (MAB) is shown in Algorithm~\ref{alg:UCBMAB}. It calculates an upper confidence bound for each action's expected reward, allowing the agent to prioritize actions with potentially higher payoffs while also exploring actions with uncertain rewards to gather more information. In our case, we initially set up action-value estimates (Q(a)) and action counts (N(a)) for all arms (a). The exploration parameter (c > 0) is chosen, and the timestep (t) is initialized as 1. Throughout our experiments, we iterate the following steps until the timestep (t) reaches the maximum limit (T). At each timestep, we select the arm (At) that maximizes the UCB value, computed as the sum of the action-value estimate (Q(a)) and an exploration term, which involves the exploration parameter (c) and the square root of the natural logarithm of the action count (N(a)). Subsequently, we execute the chosen arm and observe the resulting reward (Rt). We then update the action count (N(At)) and action-value estimate (Q(At)) for the selected arm based on the observed reward. The action count for the chosen arm is incremented by 1, and the action-value estimate is updated using the incremental update formula. This process is repeated until the maximum time (T) is reached, allowing us to efficiently explore and exploit different arms dynamically and enabling effective decision-making in uncertain environments. \\
\textbf{Complexity Analysis:} Considering the overall operations, the dominant time complexity of the algorithm is determined by the arm selection step, which is O(K). Other operations within the loop are constant time operations and do not significantly contribute to the overall time complexity. Therefore, the time complexity of the given algorithm is O(K*T), where K is the number of arms and T is the total number of timesteps the algorithm runs. \\

\begin{algorithm}
\caption{Upper Confidence Bound (UCB) Algorithm for Multi-Arm Bandit}
\begin{algorithmic}
\STATE \textbf{Initialize:} Action-value estimates $Q(a)$ and action counts $N(a)$ for all arms $a$
\STATE \textbf{Initialize:} Exploration parameter $c > 0$
\STATE \textbf{Initialize:} Timestep t as 1
\WHILE{$t \leq T$}
\STATE Choose arm $A_t$ with highest UCB value:
\STATE $A_t \leftarrow \arg\max_{a}\left(Q(a) + c \sqrt{\frac{\ln(t)}{N(a)}}\right)$
\STATE Take action $A_t$ and observe reward $R_t$
\STATE Update action count:
\STATE $N(A_t) \leftarrow N(A_t) + 1$
\STATE Update action-value estimate:
\STATE $Q(A_t) \leftarrow Q(A_t) + \frac{1}{N(A_t)}(R_t - Q(A_t))$
\ENDWHILE
\end{algorithmic}
\label{alg:UCBMAB}
\end{algorithm}

 We model a Multi-Armed Bandit for each service, where each arm represents an edge server. These arms can be 'pulled' i.e. the task can be migrated to the server, and a reward (that is the inverse of the current power consumption of the server) is received. The number of times each service has been migrated to the server is stored, such that, over time, the servers which have minimal power consumption while running the service is chosen with greater frequency.

 \subsubsection{\textcolor{black}{Proposed Model: Deep Q-Networks}}

 Deep Q-Networks is a reinforcement learning algorithm that combines Deep Neural Networks (DNN) with the Q-Learning technique \cite{DeepQLearning}. An algorithmic representation of Deep Q-Networks is shown in Algorithm~\ref{alg:DeepQL}. It enables agents to learn optimal action-selection strategies in environments with large state spaces. By approximating the action-value function using deep neural networks, Deep Q-Learning can handle complex and high-dimensional input data. Before beginning the training process, we initialise the Q-networks and learning rate ($\alpha$). We sample a state (S) from the environment and select an action (A) using a greedy strategy based on the current Q-network during each episode of the training. The following state (S') and its accompanying reward (R) are then observed from the outside world. The Q-Learning equation, which incorporates the learning rate ($\alpha$) and the discount factor ($\gamma$), is used to update the Q-values for the current state-action pair (S, A). Until convergence or a predetermined number of episodes (numEpisode), the process iterates over several episodes. By interacting with the environment, this method enables the agent to discover the best Q-values for the task at hand and gradually improve its performance. \\
\textbf{Complexity Analysis:} The time complexity for each iteration of the while loop (one episode) is O(|A|), where |A| is the number of actions. Since the while loop runs for "numEpisodes" times (N), the overall time complexity of the algorithm is O(N * |A|). \\
\begin{algorithm}
\caption{Deep Q-Learning}
\begin{algorithmic}
\STATE \textbf{Initialize : }$Q - network$
\STATE \textbf{Initialize : }learning rate $\alpha$
\WHILE {episode $<$ num\_episodes}
\STATE $S \leftarrow $ sample state from environment
\STATE $A \leftarrow \epsilon$  - greedy policy from Q network
\STATE $S’,R \leftarrow $ Next state and reward from environment
\STATE $Q(S,A) \leftarrow Q(S,A) + \alpha[R + \gamma max_{a}(Q(S',a) - Q(S,a))]$
\ENDWHILE
\end{algorithmic}
\label{alg:DeepQL}
\end{algorithm}

 We use a Deep Q-Network to determine the best server to migrate the services in the queue. The input to the Q-Network consists of a feature vector, consisting of the available CPU, available RAM, available Disk space, and the current power consumption of the edge server. The feature vector of all the edge server are concatenated and fed into the Q-Network. The Q-Network outputs a Q-value for each edge server. The edge server with the maximum Q-value is chosen for the service to be migrated to and the inverse of the power consumption is summed up and given as the reward.

 \subsubsection{\textcolor{black}{Proposed Model: Deep Q-Networks with Graphical Neural Network }}

 Graphical Neural Network (GNN) is a neural network architecture specifically designed for processing graph-structured data \cite{GraphicalNeuralNeutral}. It leverages the graph structure to capture dependencies and relationships between entities, enabling efficient representation and learning from graph data. GNN have shown promising results in tasks such as node classification, link prediction, and graph generation. An algorithmic representation of Graph Neural Network (GNN) for Message Passing is shown in Algorithm~\ref{alg:DeepQLGNN}. In our work, we employ a GNN to update node features in a given graph = (V, E). We use node features $\mathcal{E}$ and edge features $\mathbf{X}$ as input. The GNN consists of multiple layers with message aggregation and node feature update functions. For each layer, we aggregate messages mv from neighboring nodes for each node v, ($v \in \mathcal{V}$), using a message aggregation function >(v,$\mathbf{X}$). Then, we update node features $\mathbf{X}$'V using an update function UpdateNodeFeatures(v,$\mathbf{X}$,mv). This process is repeated for the specified number of layers, and the updated node features $\mathbf{X}$' are used in the next layer. The GNN captures graph information and improves node representations for downstream tasks. Experimental evaluation on benchmark datasets demonstrates its effectiveness. \\
\textbf{Complexity Analysis:} The total time complexity for each layer is dominated by the message aggregation step because it takes O(d) time per node, whereas the node feature update step takes O(1) time per node. The given algorithm consists of "numlayers" iterations, where each iteration involves both message aggregation and node feature update steps. Thus, the overall time complexity for the algorithm is: O(numlayers × |V| × d) Where:

\begin{itemize}
  \item |V| is the number of nodes in the graph.
  \item d is the average degree of nodes in the graph (assuming it's a constant).
\end{itemize}

\begin{algorithm}
\caption{Graph Neural Network (GNN) for Message Passing}
\begin{algorithmic}
\STATE \textbf{Input:} Graph $\mathcal{G} = (\mathcal{V}, \mathcal{E})$, Node Features $\mathbf{X}$, Edge Features $\mathbf{E}$
\STATE \textbf{Output:} Updated Node Features $\mathbf{X}'$

\STATE \textbf{Initialize:} GNN layers, message aggregation and update functions
\FOR{$\text{layer} = 1$ to $\text{num\_layers}$}
    \STATE \textbf{Message Aggregation:} 
    \FOR{each node $v \in \mathcal{V}$}
        \STATE $\mathbf{m}_v \leftarrow \text{AggregatesMessages}(v, \mathcal{N}(v), \mathbf{X})$
    \ENDFOR
    
    \STATE \textbf{Update Node Features:} 
    \FOR{each node $v \in \mathcal{V}$}
        \STATE $\mathbf{X}'_v \leftarrow \text{UpdateNodeFeatures}(v, \mathbf{X}, \mathbf{m}_v)$
    \ENDFOR
    
    \STATE $\mathbf{X} \leftarrow \mathbf{X}'$ \COMMENT{Update node features for the next layer}
\ENDFOR

\end{algorithmic}
\label{alg:DeepQLGNN}
\end{algorithm}

 We model the edge servers and the links between edge servers as a Graph, and feed it into a graphical neural network. Each edge server is a node with a feature vector,  consisting of the available CPU, available RAM, available Disk space, and the current power consumption of the edge server. The GNN uses message passing to aggregate information from adjacent edge servers, and finally returns a feature vector, to be fed into the Deep Q-Network.

 \subsubsection{\textcolor{black}{Proposed Model: Actor-Critic Network}}
Actor-Critic is also a reinforcement learning framework that combines elements of both value-based and policy-based methods \cite{ActorCriticReinforcementLearning}. An algorithmic representation of Actor-Critic Algorithm with Probabilistic Actions is shown in Algorithm~\ref{alg:ActorCritic}. It consists of two components: the actor, which selects actions based on a policy, and the critic, which estimates the value function or action-value function. The actor learns to improve the policy, while the critic provides feedback to the actor by estimating the quality or advantage of chosen actions. The actor network $\pi_{\theta}(a|s)$ with parameters ${\theta}$ and the critic network $V_{\phi}(s)$ with parameters $\phi$ are initialized. Additionally, learning rates $\alpha_{\theta}$ and $\alpha_{\phi}$, as well as the discount factor ${\gamma}$, are set. During the training process, episodes are executed, where each episode involves interacting with the environment and collecting states, actions, and rewards at each time step. We compute the discounted rewards-to-go for each step in the episode using the discount factor ${\gamma}$. The critic network is updated using the Advantage function, which is the difference between the discounted reward and the estimated state value from the critic network. The actor network is updated using the log-likelihood gradient, which is scaled by the Advantage function. By iteratively updating the actor and critic networks, our methodology enables the reinforcement learning system to learn effective policies for the given task. The outer loop runs for a specified number of episodes, and the inner loop runs within each episode until it is completed. In each step, the actor network samples a probabilistic action based on the current state, and the critic network estimates the state value. The algorithm then computes discounted rewards-to-go for each step in the episode and calculates the Advantage function. Using these values, it updates the critic and actor networks through gradient ascent. \\
\textbf{Complexity Analysis:} The overall time complexity of the algorithm is approximately $O(E * L^2)$, where E is the number of episodes and L is the average episode length. \\
\begin{algorithm}
\caption{Actor-Critic Algorithm with Probabilistic Actions}

\begin{algorithmic}
\STATE \textbf{Initialize:} Actor network $\pi_{\theta}(a|s)$ with parameters $\theta$, Critic network $V_{\phi}(s)$ with parameters $\phi$
\STATE \textbf{Initialize:} Learning rates $\alpha_{\theta}$, $\alpha_{\phi}$, Discount factor $\gamma$

\WHILE{episode $<$ num\_episodes}
    \STATE Initialize episode-specific lists: $\text{states} = [], \text{actions} = [], \text{rewards} = []$
    \STATE Receive initial state $s$
    \WHILE{episode not finished}
        \STATE Sample probabilistic action $a \sim \pi_{\theta}(a|s)$
        \STATE Execute action $a$, observe reward $r$ and next state $s'$
        \STATE Append $s$ to $\text{states}$, $a$ to $\text{actions}$, and $r$ to $\text{rewards}$
        \STATE $s \leftarrow s'$
    \ENDWHILE
    
    \STATE Compute the discounted rewards-to-go for each step in the episode:
    \STATE \textbf{for} $t$ from $T$ to $0$
      
       \STATE \COMMENT{where $T$ is the last time step in the episode}
       \STATE $G_t = r_t + \gamma . G_{t+1} Update G_t with discount factor $
    \STATE \textbf{End for}
    
    \FOR{$t$ from $0$ to $T$}
        \STATE Compute the Advantage function:
        \STATE $A_t = G_t - V_{\phi}(s_t)$ \COMMENT{Using the critic network to estimate the state value}
        \STATE Update the Critic network:
        \STATE $\phi \leftarrow \phi + \alpha_{\phi} \cdot \nabla_{\phi} V_{\phi}(s_t) \cdot A_t$
        \STATE Update the Actor network using the log-likelihood gradient:
        \STATE $\theta \leftarrow \theta + \alpha_{\theta} \cdot \nabla_{\theta} \log \pi_{\theta}(a_t|s_t) \cdot A_t$
    \ENDFOR
\ENDWHILE
\end{algorithmic}
\label{alg:ActorCritic}
\end{algorithm}
 Each edge server's CPU demand, memory demand, hard disk demand, and power consumption are concatenated into a feature vector and fed into an actor deep network, as well as a critic deep network. The actor outputs the probability of choosing each server, while the critic outputs the 'Q - value', the value of being in the state as given by the probability vector. The probability vector is sampled to choose the edge server to place the service in while the Q-value, along with the reward, is used for updation.

 \subsection{\textcolor{black}{Communication among entities}}

\begin{figure}[t]
    \centering
    \includegraphics[width=1\linewidth]{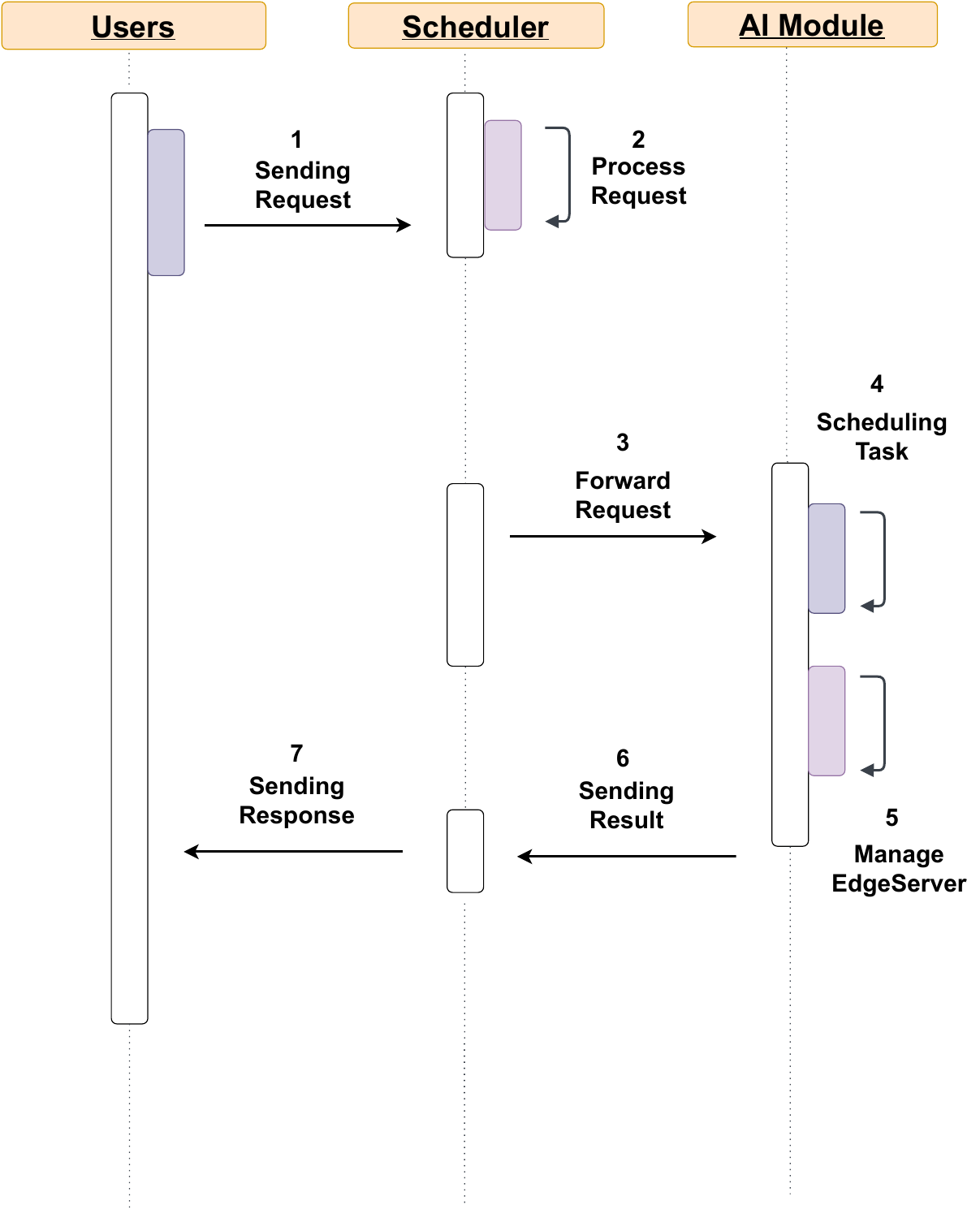}
    \caption{\textcolor{black}{Simulation data flow among users, scheduler and AI module}}
    \label{fig:sequence}
\end{figure}

\textcolor{black}{Fig \ref{fig:sequence} illustrates the simulation data flow among users, scheduler and AI module in the EdgeAISim. The scheduler receives requests from the user, processes them, and then chooses the appropriate AI module. Then the edge server is managed and the task is scheduled by the AI Module. Results from completed tasks are sent to the scheduler, which subsequently sends them to the user.}

\section{Experiments and Evaluation}

In our research, we employed a comprehensive approach to measure the power consumption of individual edge servers and the total power consumed at each time step in the edge computing system. This allowed us to gain valuable insights into the energy usage patterns of the entire system. By closely monitoring the power consumption of each edge server, we could pinpoint variations in energy usage across different resources, leading to a deeper understanding of the system's overall efficiency. Simultaneously, tracking the total power consumption over time helped us analyze energy consumption trends and assess the system's performance in terms of power management and optimization strategies. This data-driven analysis formed the foundation for making informed decisions to improve energy efficiency and optimize the edge computing infrastructure.

\subsection{Experimental Setup}\label{sec:Setup}

\textcolor{black}{We utilized an Intel HexaCore i7 - 8750 H processor, which uses coffee lake architecture \cite{abel2019uops}, with an NVIDIA 1050 Ti GPU with 16 GB of RAM and 4 GB Video RAM (VRAM), using Pascal architecture \cite{lombardi2019pascal} on the Linux Operating System to run our simulations. We used the PyTorch library in order to implement the deep reinforcement learning algorithms. Table \ref{Hyperparameters} shows the hyperparameters and their corresponding values, which are used for these experiments. }

\subsection{Workloads/Dataset}\label{sec:Workload}

\textcolor{black}{We considered a scenario where there are 6 edge servers with the CPU, memory and disk capacities as specified in Table \ref{Server} and CPU, Memory and disk demands from pre-existing workloads as specified in Table \ref{ServerDemands}.}

\begin{table}[htbp]

    \centering
    \caption{Server Specifications}
    \label{tab:specs}
    \begin{tabular}{|l|c|c|c|}
        \hline
        \textbf{Server} & \textbf{CPU} & \textbf{Memory} & \textbf{Disk} \\
        \hline
        Edgeserver1 & 8 & 16384 & 131072 \\
        Edgeserver2 & 8 & 16384 & 131072 \\
        Edgeserver3 & 8 & 8192  & 131072 \\
        Edgeserver4 & 8 & 8192  & 131072 \\
        Edgeserver5 & 12 & 16384 & 131072 \\
        Edgeserver6 & 12 & 16384 & 131072 \\
        \hline
    \end{tabular}
    \label{Server}
\end{table}

\begin{table}[htbp]
    \centering
    \caption{Server Demands}
    \label{tab:demands}
    \begin{tabular}{|l|p{1.2cm}|p{1.2cm}|p{1.0cm}|} % Customize the column width as needed
        \hline
        \textbf{Server} & \textbf{CPU Demand} & \textbf{Memory Demand} & \textbf{Disk Demand} \\
        \hline
        Edgeserver1 & 0 & 0 & 0 \\
        Edgeserver2 & 0 & 0 & 0 \\
        Edgeserver3 & 0 & 0 & 0 \\
        Edgeserver4 & 0 & 0 & 0 \\
        Edgeserver5 & 1 & 1024 & 1017 \\
        Edgeserver6 & 0 & 0 & 0 \\
        \hline
    \end{tabular}
    \label{ServerDemands}
\end{table}

\begin{table}[htbp]
    \centering
    \caption{\textcolor{black}{Hyperparameters and Values}}
    \label{tab:demands}
    \begin{tabular}{|l|p{2.5cm}|p{2.3cm}} % Customize the column width as needed
        \hline
        \textbf{Hyperparameter} & \textbf{Hyperparameter value} \\
        \hline
        $\alpha$ (learning rate) & 0.05  \\
        $\epsilon$ (exploration factor) & 1 \\
        $\gamma$ (Future reward weight factor) & 0.9 \\
        $\epsilon$ decay (How much epsilon decreases) & 0.997 \\
        Dropout probability in NN & 0.5 \\
        \hline
    \end{tabular}
    \label{Hyperparameters}
\end{table}

\subsection{Experimental Results}\label{sec:Results}
This section presents the experimental results for different AI models (RL algorithms) such as Multi-Armed Bandit with Upper Confidence Bound, Deep Q-Networks, Deep Q-Networks with Graphical Neural Network, and Actor-Critic Network and baseline model using Worst-Fit Migration Algorithm.

\subsubsection{Baseline Model: Worst-Fit Migration Algorithm}\label{sec:Baseline}
\begin{figure}[htbp]
\centerline{\includegraphics[width=0.8\linewidth]{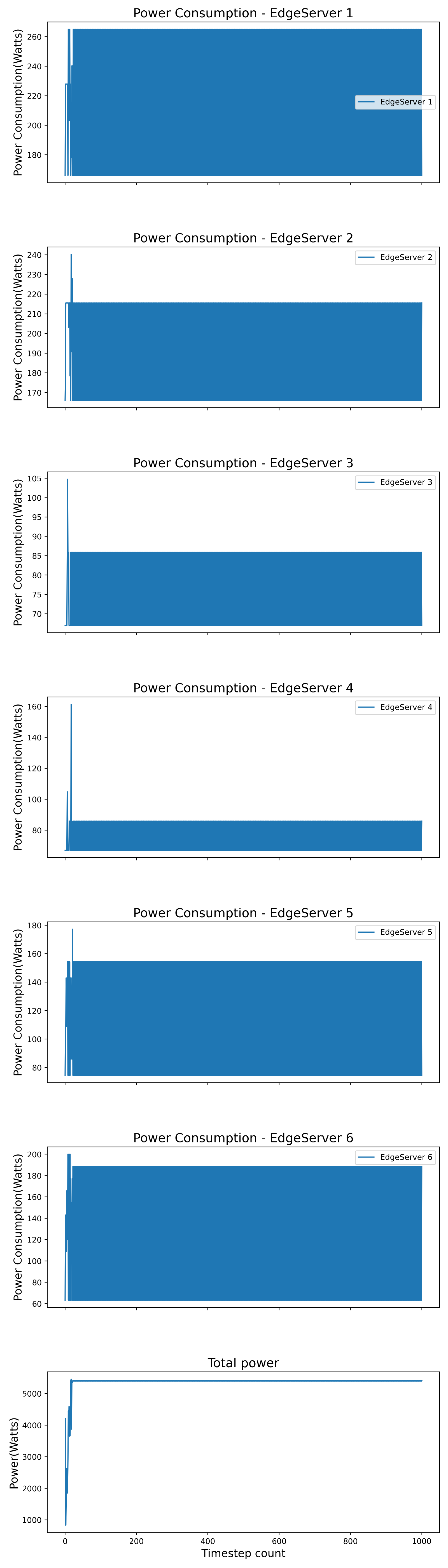}}
\caption{\textcolor{black}{Power consumption for 6 servers, with total power consumption for the worst-fit migration algorithm}}
\label{fig5}
\end{figure}

\textcolor{black}{Figure \ref{fig5} shows the power consumption for 6 servers, with total power consumption for the baseline which is worst-fit migration algorithm based resource management policy \cite{lai2020cost, xu2022dynamic}. Total power consumption in the baseline worst fit migration rapidly increases from the start till it reaches about 6000 Watt (W). This is due to the algorithm allocating tasks to the server with the most resources available, without regard for power consumption.}

\subsubsection{Proposed Model: Multi-Armed Bandit (MAB) with Upper Confidence Bound (UCB)}
\begin{figure}[htbp]
\centerline{\includegraphics[width=0.8\linewidth]{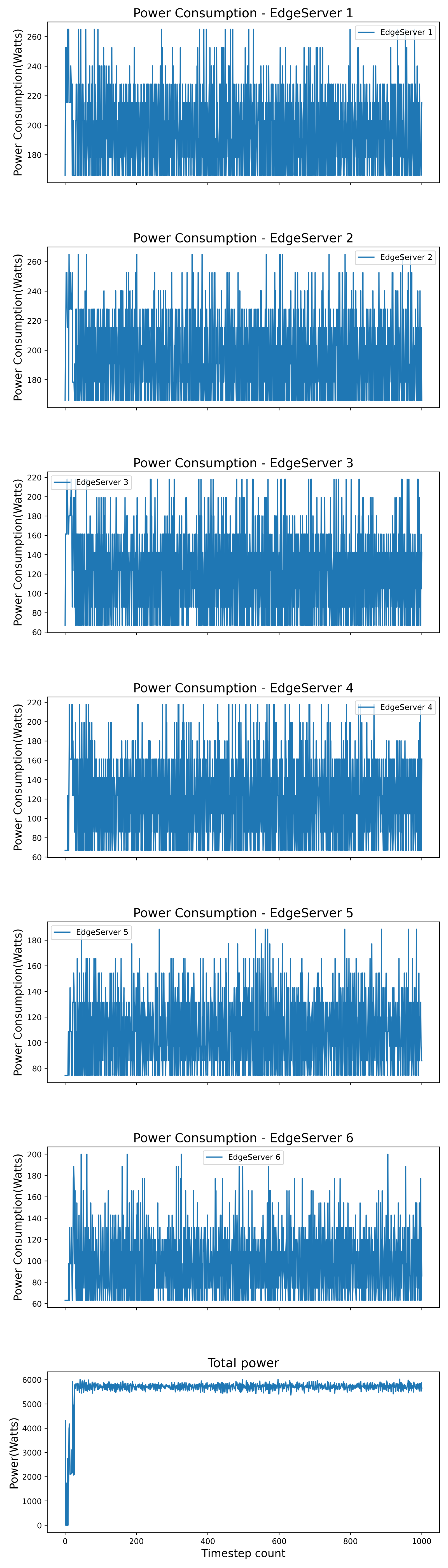}}
\caption{\textcolor{black}{Power consumption for 6 servers, with total power consumption for MAB with UCB.}}
\label{fig6}
\end{figure}

Figure \ref{fig6} shows the power consumption for 6 servers, with total power consumption for MAB with UCB. The observed stagnation of power consumption at around 6000 W, which is comparable to the baseline, presents significant implications for the multi-arm bandit algorithm that was specifically designed to minimize power consumption through task migration. The algorithm's primary objective is to dynamically allocate computational tasks among multiple resources to achieve optimal power efficiency. However, the plateau in power consumption suggests that the current approach might have reached its limits in terms of further power reduction. This raises critical considerations for the algorithm's effectiveness and the potential need for reassessing its underlying assumptions or exploring alternative strategies. \textcolor{black}{This stagnation shows a 0 or negligible percentage of improvement.}

\subsubsection{Proposed Model: Deep Q-Networks}
\begin{figure}[htbp]
\centerline{\includegraphics[width=0.8\linewidth]{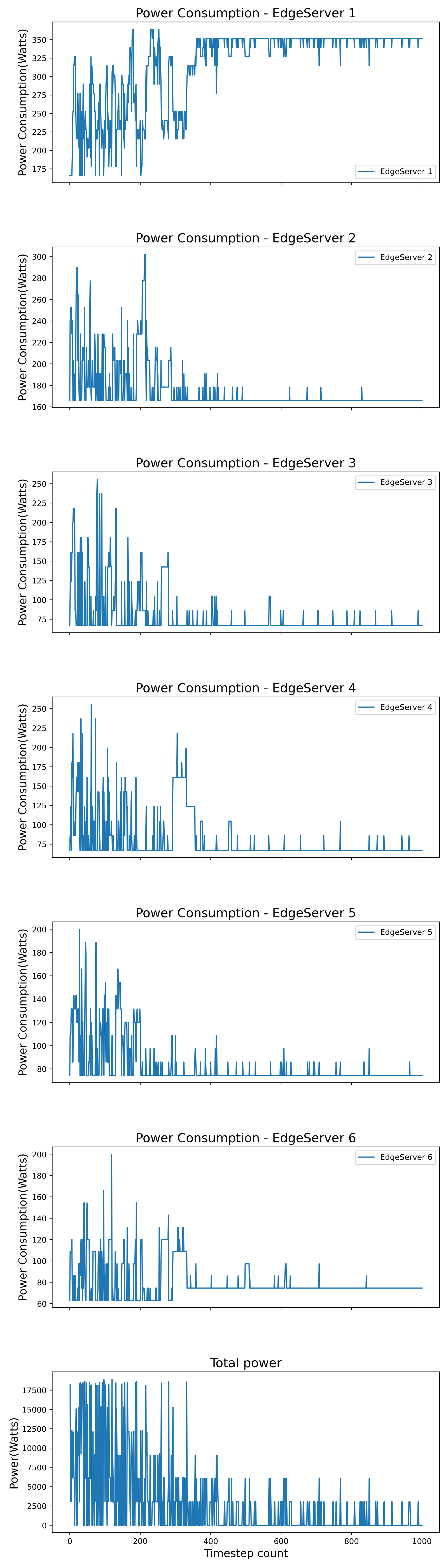}}
\caption{\textcolor{black}{Power consumption for 6 servers, with total power consumption for Deep Q-Networks.}}
\label{fig7}
\end{figure}

Figure \ref{fig7} shows the power consumption for 6 servers, with total power consumption for Deep Q-Networks. The observed minimization of total power consumption is a significant achievement for the Deep Q-Network algorithm designed to optimize power usage through task migration. The algorithm's ability to more prominently utilize the servers with lower power consumption, such as Edgeserver-1, suggests that it has successfully learned to make intelligent decisions regarding task allocation. By leveraging the power of deep reinforcement learning, the algorithm demonstrates its capacity to adaptively distribute tasks across servers, ultimately leading to reduced energy consumption. \textcolor{black}{There is an initial high allocation by Q-Learning, which leads to a power consumption of 17500W, which however drops significantly to around 2500W. This is an improvement of about 58\%, as compared to 6000 W in baseline.}

\subsubsection{Proposed Model: Deep Q-Networks with Graphical Neural Network (GNN)}
\begin{figure}[htbp]
\centerline{\includegraphics[width=0.8\linewidth]{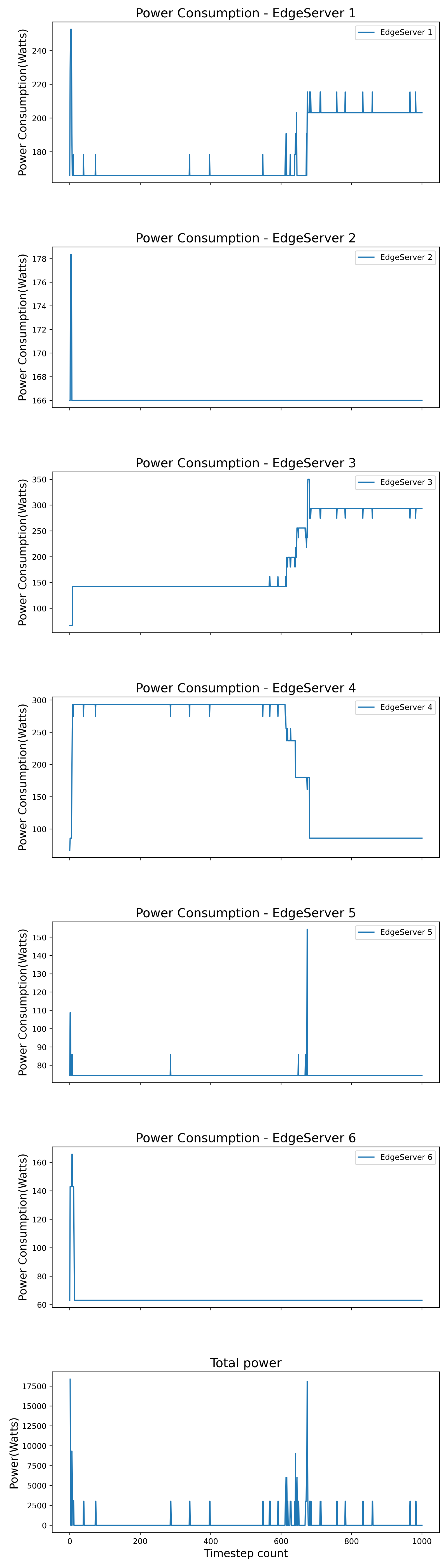}}
\caption{Power consumption for 6 servers, with total power consumption for Deep Q-Networks with Graphical Neural Network (GNN).}
\label{fig8} 
\end{figure} 

Figure \ref{fig8} shows the power consumption for 6 servers, with total power consumption for Deep Q-Networks with GNN. The successful minimization of total power consumption and the clear trend of favoring servers with lower power consumption (EdgeServer-3 and EdgeServer-4) are remarkable outcomes for the Deep Q-Networks algorithm enhanced with GNN to optimize power usage through task migration. The incorporation of GNN enables the algorithm to effectively capture and exploit the underlying relationships between edge servers, facilitating more informed and strategic task allocation decisions. By leveraging the power of GNN, the algorithm can better identify the most energy-efficient servers for task offloading, resulting in significant reductions in overall power consumption. \textcolor{black}{Similar to the previous experiment, power consumption by Deep Q-Networks with GNN is initially high due to a large allocation of 17500W; however, it drops dramatically to roughly 2500W. When compared to a baseline output of 6000 W, this is an increase in efficiency of almost 58\%.}

\subsubsection{Proposed Model: Actor-Critic Network}
\begin{figure}[htbp]
\centerline{\includegraphics[width=0.8\linewidth]{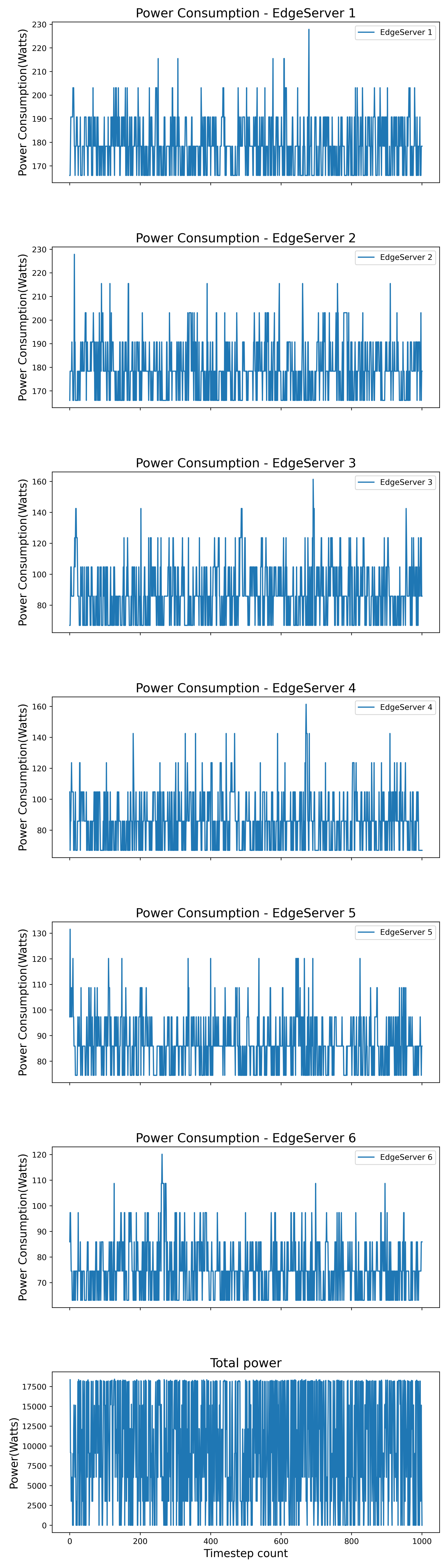}}
\caption{Power consumption for 6 servers, with total power consumption for a probabilistic actor-critic network.}
\label{fig9}
\end{figure}

Figure \ref{fig9} shows the power consumption for 6 servers, with total power consumption for a probabilistic Actor-Critic Network. The observed stagnation at a high power consumption level is a concerning outcome for the Actor-Critic network algorithm designed to minimize power consumption through task migration. Despite its potential for continuous action space exploration, the algorithm's inability to effectively learn indicates challenges in capturing the complex relationships between tasks and edge servers. As a result, it struggles to make informed decisions on task allocation, leading to sub-optimal power consumption levels. \textcolor{black}{Similar to the MAB with UCB, this stagnation and wild fluctuation show a 0 or negligible percentage of improvement using the Actor-Critic Network.}

\subsection{Performance Comparison of EdgeAISim with Baseline Model}

\textcolor{black}{Table \ref{table:comparisntable} shows the comparison of AI models of EdgeAISim with baseline \cite{lai2020cost, xu2022dynamic} in terms energy consumed on different servers. Notably, two AI models within EdgeAiSim, Q-Learning and Q-Learning with GNN, demonstrate reduced overall power consumption in contrast to the baseline worst fit migration algorithm. This positive result can be credited to the integration of algorithmic improvements and the adoption of efficient application strategies embedded within these two models, which leverage reinforcement learning and graph-based machine learning methodologies. As a result, EdgeAISim has effectively presented substantial reductions in power usage within the realm of edge computing, thereby providing valuable insights and serving as a source of inspiration for the implementation of sustainable edge computing solutions on a broader scale.}

\begin{table*}[h]
\centering  
\caption{\textcolor{black}{Comparison of EdgeAISim with baseline in terms of Energy Consumption for different servers}} 
\label{table:comparisntable} 
\begin{tabular}{@{}|c|c|c|c|c|c|c|c|c|@{}}
\toprule
\textbf{Works} &
  \textbf{Models} &
  \textbf{\begin{tabular}[c]{@{}c@{}}Edge \\ Server 1  \\  (W)\end{tabular}} &
  \textbf{\begin{tabular}[c]{@{}c@{}}Edge \\ Server 2  \\ (W)\end{tabular}} &
  \textbf{\begin{tabular}[c]{@{}c@{}}Edge \\ Server 3 \\  (W)\end{tabular}} &
  \textbf{\begin{tabular}[c]{@{}c@{}}Edge \\ Server 4  \\ (W)\end{tabular}} &
  \textbf{\begin{tabular}[c]{@{}c@{}}Edge \\ Server 5  \\ (W)\end{tabular}} &
  \textbf{\begin{tabular}[c]{@{}c@{}}Edge \\ Server 6  \\ (W)\end{tabular}} &
  \textbf{\begin{tabular}[c]{@{}c@{}}Total \\ Power  \\ (W)\end{tabular}} \\ \midrule
\multirow{4}{*}{EdgeAISim} & \begin{tabular}[c]{@{}c@{}}Q-Learning  \\ with GNN\end{tabular} &  181.4  & 166.24 & 184.5  &  200 & 81 & 63 & 1000   \\ \cmidrule(l){2-9} 
                           & Q-Learning                                                 &  307  & 180 & 85.9  &  91.25 & 90.5 & 82.4 & 4875 \\ \cmidrule(l){2-9} 
                           & Actor-Critic                                             & 236.5 & 188.5 & 94.4 &96  & 81.9 & 79.6 & 12850 \\ \cmidrule(l){2-9} 
                           & MAB                                                        & 217.5 & 227  & 163 & 161.5 & 128.6 & 129 & 5600 \\ \midrule
Baseline \cite{lai2020cost, xu2022dynamic}                  & Worst fit                                                  & 220 & 192.57 & 77.85 & 80.85 & 118.5 & 125  & 5431 \\ \bottomrule
\end{tabular}
\end{table*}

\section{Conclusions and Future Work}

\textcolor{black}{The surge in IoT-based applications has led to an unprecedented volume of data, posing significant challenges for traditional cloud computing due to issues such as latency, network traffic limitations, and security concerns. Edge computing offers a promising solution by bringing processing power and storage closer to the network's edge, reducing latency and alleviating network traffic issues. However, edge devices possess limited resources, necessitating efficient resource management with a focus on optimizing power consumption. In this paper, we proposed EdgeAISim, a novel framework that leverages AI models (especially reinforcement learning algorithms) and task migration strategies to address this critical concern in edge computing. By integrating advanced AI models such as Multi-Armed Bandit with Upper Confidence Bound, Deep Q-Networks, Deep Q-Networks with Graphical Neural Network, and Actor-Critic Network into the existing EdgeSimPy framework, we have successfully demonstrated the capability of EdgeAISim to significantly reduce power consumption in edge computing environments. In summary, our work showcases the potential of EdgeAISim as a powerful tool for enhancing the sustainability and efficiency of edge computing, offering tangible solutions to the pressing challenges posed by the ever-expanding IoT landscape.}

\subsection{Promising Future Directions}

\textcolor{black}{
While this work primarily focuses on mitigating energy consumption challenges in edge computing by proposing the basic version of EdgeAISim, it provides a solid foundation for several promising avenues of research and development. These future directions include:}

\subsubsection{IoT Applications}
\textcolor{black}{
 Exploring how EdgeAISim can be tailored to specific IoT use cases will be valuable. Different IoT applications, from smart cities to healthcare, have unique demands, and adapting the framework to these requirements will be essential \cite{edgeai}.}
 \subsubsection{6G Networks and Beyond}
 \textcolor{black}{
As next-generation networks like 6G networks become more prevalent, investigating how EdgeAISim can leverage the increased connectivity and bandwidth to further improve edge computing performance and efficiency is vital \cite{gill2022ai}.}
 \subsubsection{Security and Privacy}
 \textcolor{black}{
Expanding the EdgeAISim framework to address security and privacy concerns in edge computing environments is essential. Developing mechanisms to safeguard sensitive data and ensure secure task migration will be crucial in future deployments \cite{gill2022ai}.}
 \subsubsection{Multi-objective Optimization}
 \textcolor{black}{
Moving beyond energy consumption, researchers can explore multi-objective optimization using EdgeAISim that considers a balance between energy efficiency, QoS, and security, among other factors, to provide a holistic solution \cite{edgeai}.}
 \subsubsection{Edge Device Heterogeneity}
 \textcolor{black}{
Investigating how EdgeAISim can adapt to diverse edge device capabilities and resource constraints will be necessary to accommodate the increasing variety of devices in edge environments \cite{iftikhar2022ai}.}
 \subsubsection{Real-world Deployment and Validation}
 \textcolor{black}{
Finally, conducting real-world deployments and validations of EdgeAISim in various edge computing scenarios and industries will be a significant step toward practical implementation \cite{iftikhar2022ai}.}
\par 
\textcolor{black}{
Incorporating these aspects into the development and enhancement of EdgeAISim will ensure its relevance and effectiveness in addressing the evolving challenges and opportunities in the dynamic field of edge computing.}

\section*{Software Availability}
We released EdgeAISim available for free as open source. All code, datasets, and result reproducibility scripts are publicly available and can be accessed from GitHub:
\url{https://github.com/MuhammedGolec/EdgeAISim}

\section*{Acknowledgements}
The research investigation was carried out during the internships undertaken by Aadharsh Roshan Nandhakumar, Ayush Baranwal, and Priyanshukumar Choudhary at Queen Mary University of London, UK. Muhammed Golec would express his thanks to the Ministry of Education of the Turkish Republic, for their support and funding.

\section*{Conflict of Interest}
On behalf of all authors, the corresponding author states that there is no conflict of interest.

\section*{Credit authorship contribution statement}
\textbf{Aadharsh Roshan Nandhakumar} Conceptualization, Data curation, Investigation, Methodology, Software, Visualization, Validation, Formal analysis, Writing - original draft. 
\textbf{Ayush Baranwal:} Visualization, Validation, Formal analysis, Writing - original draft. 
\textbf{Priyanshukumar Choudhary:} Conceptualization, Data curation, Investigation, Methodology, Writing - original draft.
\textbf{Muhammed Golec:} Conceptualization, Visualization, Validation, Formal analysis, Writing - original draft and Supervision.  
\textbf{Sukhpal Singh Gill:} Conceptualization, Data curation, Investigation, Methodology, Validation, Formal analysis, Writing - original draft and Supervision.

\bibliographystyle{ieeetr}

\bibliography{cas-refs}

\end{document}